\documentclass[
superscriptaddress,
preprint,
nofootinbib,
amsmath,amssymb,
aps,
prd,
floatfix,
]{revtex4-2}
\usepackage[dvipsnames]{xcolor}
\usepackage{graphicx}
\usepackage{bm}
\usepackage[colorlinks=true,hyperfootnotes=true,breaklinks=true,citecolor=PineGreen,linkcolor=PineGreen]{hyperref}

\usepackage{physics}

\begin{document}
\begin{titlepage}
\begin{center}

{\LARGE \textbf{Quantum Field Theory in Curved Spacetime Approach to the Backreaction of Dynamical Casimir Effect}}
\vspace{1.5cm}

\begin{figure}[htbp]
    \centering
    \includegraphics[width=0.35\textwidth]{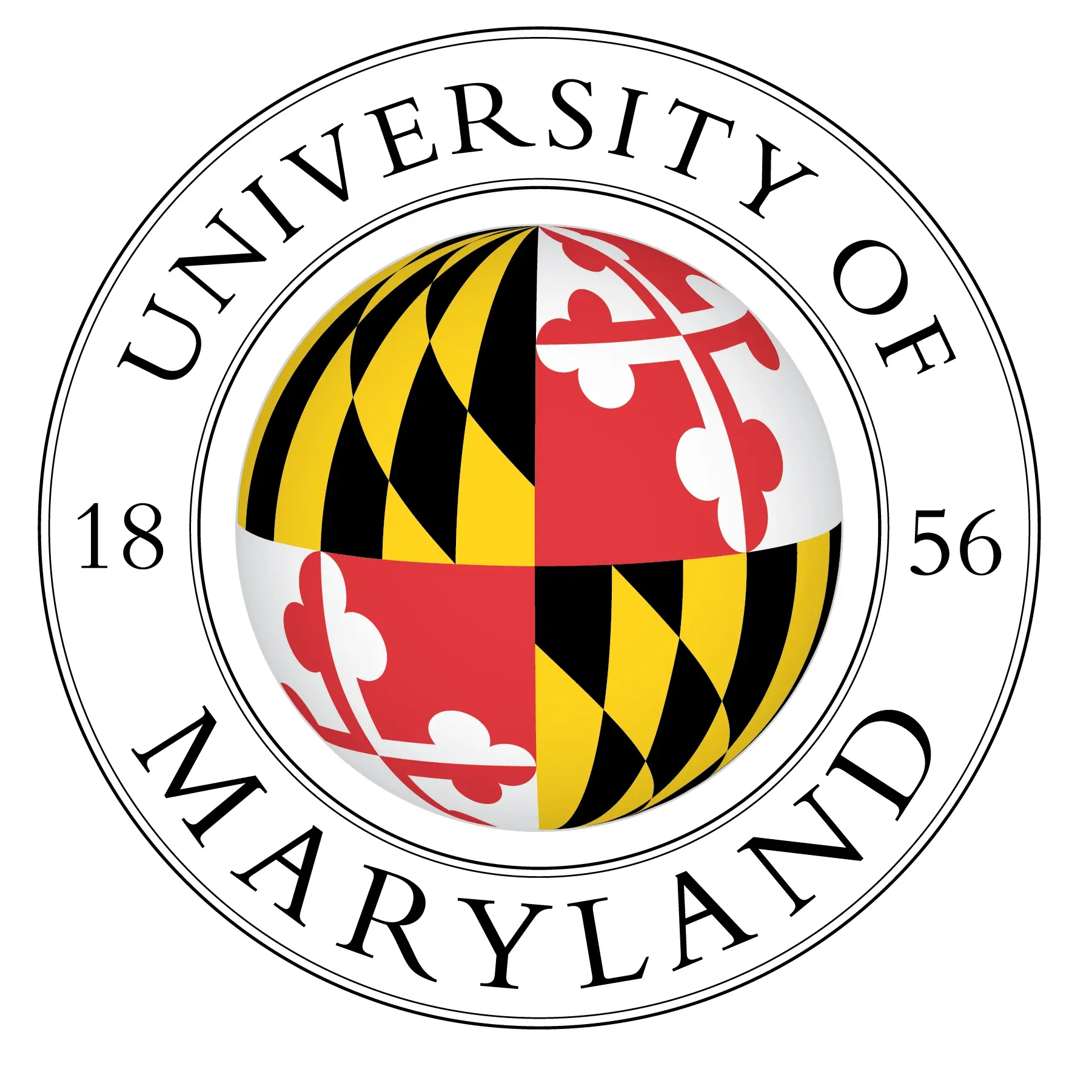}
\end{figure}
\vspace{1cm}
\textsc{\LARGE Yu-Cun Xie}\\[0.7cm]
{\Large Department of Physics}\\[0.7cm]
{\Large University of Maryland, College Park} \\[1cm]
{\Large Undergraduate Honor Thesis} \\[0.7cm]
{\Large \textit{Advisor:} Bei-Lok Hu Ph.D.} \\[0.7cm]

\vspace{3cm}

\large \text{March 2024}
\thispagestyle{empty}
\end{center}
\end{titlepage}

\newpage
\thispagestyle{empty}
\begin{center}
    {\large{\textbf{Acknowledgments}}}
\end{center}
\vspace{1cm}

During my time as an undergraduate, I was fortunate to receive a wealth of support and assistance from the University of Maryland, which was instrumental in the success of my research endeavors. I am particularly grateful to the Department of Physics for the unwavering support I received from numerous faculty and staff members. I would thank Mr. Logan Anbinder and Ms. Stephanie Williams for being my academic advisors and providing much help in my study.
I extend my heartfelt thanks to Dr. Steven Anlage and Dr. Gretchen Campbell for serving on my committee and for dedicating their valuable time to assist me with my thesis and defense.

I must express my special gratitude to Dr. Ted Jacobson, whose exceptional teaching and passionate commitment to both research and education have deeply inspired me, enhancing my understanding of many physics concepts.

Above all, my deepest appreciation goes to my research advisor Dr. Bei-Lok Hu. His mentorship and profound knowledge have been my guiding light over the past two years, and I sincerely doubt I could have navigated this journey without his sage advice.

Finally, I cannot forget the emotional backbone provided by my family throughout this period. Your unwavering support has been my anchor. Thank you for always being there for me.
\newpage
\thispagestyle{empty}
\begin{center}
    {\large{\textbf{Table of Contents}}}
\end{center}
\thispagestyle{empty}
\tableofcontents
\newpage

\thispagestyle{empty}
\begin{center}
    {\large{\textbf{Abstract}}}
\end{center}
\thispagestyle{empty}
\vspace{1cm}

In this thesis, we investigate the dynamical Casimir effect, the creation of particles from vacuum by dynamical boundary conditions or dynamical background, and its backreaction to the motion of the boundary. The backreaction of particle creation to the boundary motion is studied using quantum field theory in curved spacetime technique, in 1+1 dimension and 3+1 dimension. 
The relevant quantities in these quantum field processes are carefully analyzed, including regularization of the UV and IR divergent of vacuum energy, and estimation of classical backreaction effects like radiation pressure. We recovered the qualitative result of backreaction in 1+1 dimensions. In the 3+1 dimension, we find that the backreaction tends to slow down the system to suppress the further particle creation, similar to the case of cosmological particle creation. 
\newpage

\begin{center}
    {\large{\textbf{Introduction}}}
\end{center}
\thispagestyle{empty}
\vspace{1cm}
In the last two decades, researchers in analog gravity have skillfully designed many laboratory-accessible experiments to test some important effects in semiclassical gravity, with fluids analog to Hawking-Unruh radiation being a prominent example \cite{Hawana}. Another well-known effect in semiclassical gravity is the cosmological particle creation, an effect that vacuum fluctuations of the quantum field are parametrically amplified by an expanding universe \cite{Par69,Zel70}. 

At about the same time that cosmological particle creation was discovered, G.T. Moore used a simplified one-dimensional model to show that the motion of boundaries of the one-dimensional cavity could result in the generation of photons from the vacuum \cite{DCE}. The effect was then investigated by DeWitt \cite{DEWITT1975295}, Fulling and Davies \cite{DavFul,DavFul+}, and eventually being named as dynamical Casimir effect (DCE) \cite{DCEname}. Although this effect hasn't been directly observed, it has been observed in a superconducting circuit in 2011 \cite{wilson2011observation}.

The dynamical Casimir effect shares the same underlying physical mechanism with the cosmological particle creation, that the parametric amplification induced by moving boundary. Although there are many articles concerning the dynamical Casimir effect, using simple models like two parallel conducting plates with time-varying separation \cite{ParallelPlateDCE1,ParallelPlateDCE2}, the backreaction effect remains puzzling. Solving the backreaction requires the trajectory of the mirror dynamically determined by the equation of motion, but backreaction terms in the equation of motion can only be solved under a given mirror trajectory. The backreaction of cosmological particle creation at the Planck time has been shown to play a significant role in the isotropization and homogenization of the early universe \cite{HuPar78,HarHu79,HarHu80,CalHu87}, therefore it is natural to investigate the backreaction of DCE.

The backreaction problem of DCE can be simplified if we can approximate the space between the mirror/inside the cavity to be homogeneous. This approximation can be achieved if the boundary motion is much slower than the speed of light, and neglects the initial transient state. In this case, the field mode inside the boundary will be able to adjust itself to remain overall homogeneous. If the space inside the boundary is homogeneous, one can construct the quantum field similar to some dynamical spacetime model. This approximation also neglected the force from the field outside the cavity. However, it has been shown in \cite{DODONOV1989511} that the force from the field outside the cavity is a higher-order term, and it is suppressed by the $c^3$.

All of the topics discussed in this thesis are related to \cite{Xie:DCE,Xie:DVC}. 

In Chapter \ref{QFTCS}, a brief overview of quantum field theory (QFTCS) in curved spacetime is given, with focusing on quantum field construction in cosmological spacetime and cosmological particle creation. 

In Chapter \ref{2d}, we investigate a one-dimensional simplified model, calculate the backreaction based on the assumption of homogeneity of space, and show the assumption of homogeneity gives a good approximation in the adiabatic limit. We also numerically solved the mirror equation of motion, which shows that, in two-dimension, the backreaction is always in the same direction as static Casimir force, and does not provide a damping behavior. 

In Chapter \ref{4d}, we investigate a three-dimensional model, with a rectangular box boundary and allow one side to move. With the assumption of homogeneity, the quantum field can be constructed the same as an anisotropic expanding universe, i.e. Bianchi type-I spacetime. We present a careful analysis of the stress-energy tensor and use it as the source deriving the mirror's equation of motion. We present the numerical solution of the mirror's equation of motion, which shows the mirror motion gets damped no matter what direction it is moving, similar to the backreaction of cosmological particle creation.

\newpage
\section{Quantume Field Theory in Curved Spacetime: an Overview }\label{QFTCS}
\pagenumbering{arabic}
In the Minkowski spacetime, we have a unique set of bases that are invariant under the action of the Lorentz group, which allows us to construct the Lorentz invariant vacuum state and concept of particle. However, in general spacetime, global Lorentz symmetry no longer exists, and we do not have a preferred vacuum state \& definition of particles. The consequence is that the concepts of vacuum and particle are ill-defined. Suppose the gravitational background is time-dependent, as an expanding universe. In that case, the time-translational symmetry is broken, and the field could absorb energy from the background gravitational field, which is related to cosmological particle creation. We follow the sign convention in \cite{BirDav}, which is $(-,-,-)$ convention in the terminology of Misner, Thorne \& Wheeler \cite{Misner:1973prb}. The treatment in this chapter assumes all the spacetime is globally hyperbolic.

\subsection{Construct QFT in Curved Spacetime}
In QFTCS, we usually study scalar fields with different coupling for simplicity. More complicated fields usually behave similarly to scalar fields, e.g., each polarization of the Electromagnetic field can be described by a conformal coupled scalar field (the meaning of conformal couple will be defined later). 
We start with the action of a scalar field in flat spacetime,
\begin{eqnarray}
    S = \int d^4 x\;\frac{1}{2}(\eta^{\mu\nu} \phi_{,\mu} \phi_{,\nu}- m^2 \phi^2).
\end{eqnarray}
To obtain the curved spacetime expression, we replace everything for flat spacetime with their curved spacetime correspondence, and introduce a coupling term to scalar curvature,
\begin{equation}
  S = \int d^4 x\;\frac{1}{2}\sqrt{-g}(g^{\mu\nu} \phi_{;\mu} \phi_{;\nu}- m^2 \phi^2-\xi R \phi^2),
\end{equation}
where the \(R\) is the Ricci scalar and \(\xi\) is the coupling constant. 

We usually have two favorite choices of coupling constant;
\textbf{minimal coupling} simply set \(\xi=0\) which turn off the coupling,
while \textbf{conformal coupling} sets 
\begin{eqnarray}
    \xi = \frac{(n-2)}{4(n-1)},
\end{eqnarray}
which makes the action invariant under conformal transformation when \(m=0\). Each polarization of the Electromagnetic field can be described by a conformal coupled scalar field; each polarization of the graviton field can be described by a minimal coupled scalar field \cite{HuPar77}. 

The equation of motion (Klein-Gordon equation) can obtained by stationary action principle,
\begin{equation}\label{kg}
  \left[\Box+m^2+\xi R\right]\hat{\phi}=0.
\end{equation}
The inner product of the scalar fields in curved spacetime is defined as (sometimes referred to as Klein-Gordon inner product)
\begin{eqnarray}
    (\phi_1,\phi_2)=-i\int_{\Sigma}\dd^{n-1}x\;\sqrt{\gamma}n^{\mu}\left[\phi_1\nabla_\mu\phi^*_2-\phi_2^*\nabla_\mu\phi_1\right],
\end{eqnarray}
where \(\gamma\equiv \det(\gamma_{ij})\) is the determinant of induced metric of a constant time spacelike hypersurface \(\Sigma\), and \(n^{\mu}\) is the unit normal vector of \(\Sigma\). One can prove this inner product is independent of the foliation of the hypersurface by using the generalized Stokes theorem. 

We can proceed with the canonical quantization similar to that in the flat spacetime. First, we have the conjugate momentum as
\begin{equation}
  \pi=\frac{\partial \mathcal{L}}{\partial(\nabla_{0}\phi)}=\sqrt{-g}\nabla_{0}\phi,
\end{equation}
and we impose the canonical commutation relation of the field operator in equal time hypersurface;
\begin{equation}
\begin{aligned}\label{cccr}
    \bigl[\hat{\phi}(t,\mathbf{x}),\hat{\phi}(t,\mathbf{x'})\bigr] & =0 ,                                                        \\
    \bigl[\hat{\pi}(t,\mathbf{x}),\hat{\pi}(t,\mathbf{x'})\bigr]   & =0 ,                                                        \\
    \bigl[\hat{\phi}(t,\mathbf{x}),\hat{\pi}(t,\mathbf{x'})\bigr]  & =\frac{i}{\sqrt{-g}}\delta^{(n-1)}(\mathbf{x}-\mathbf{x'}).
\end{aligned}
\end{equation}
There exists a complete orthonormal set of modes, satisfying
\begin{equation}
\begin{aligned}
  \left(f_{i},f_{j}\right)     & =\delta_{ij},   \\
  \left(f_{i},f^*_{j}\right)   & =0      ,       \\
  \left(f^*_{i},f^*_{j}\right) & =-\delta_{ij}.
\end{aligned}
\end{equation}
Where the index \(ij\) represents some quantities that label the modes.
Because the set of modes is complete, we can expand our field as
\begin{eqnarray}\label{cme}
    \hat{\phi}=\sum_i\left[{}^{(1)}\hat{a}_i f_i+{}^{(1)}\hat{a}_i^\dagger f_i^*\right].
\end{eqnarray}
The coefficients of mode expansion have commutation relations;
\begin{equation}
\begin{aligned}
\bigl[{}^{(1)}\hat{a}_{i},{}^{(1)}\hat{a}_{j}\bigr]                     & =0 ,          \\
  \bigl[{}^{(1)}\hat{a}^{\dagger}_{i},{}^{(1)}\hat{a}^{\dagger}_{j}\bigr] & =0  ,         \\
  \bigl[{}^{(1)}\hat{a}_{i},{}^{(1)}\hat{a}^{\dagger}_{j}\bigr]           & =\delta_{ij}.
\end{aligned}
\end{equation}
We interpret those coefficients as creation and annihilation operator, which defines a vacuum state,
\begin{eqnarray}
    \forall\;i,\ {}^{(1)}\hat{a}_i\ket{0_f}=0.
\end{eqnarray}
From this vacuum state, we can construct the Fock space. A state with \(n_i\) excitations could be create by repeated action of \({}^{(1)}\hat{a}^\dagger_i\),
\begin{eqnarray}
  \ket{n_i}=\frac{1}{\sqrt{n_i!}}\left({}^{(1)}\hat{a}_i^{\dagger}\ket{0_f}\right),
\end{eqnarray}
and the number operator is defined as 
\begin{eqnarray}
    \hat{n}_{f_i}={}^{(1)}\hat{a}_i^\dagger {}^{(1)}\hat{a}_i.
\end{eqnarray}
All the above constructions of Fock space and vacuum state are built for the set of modes \(f_i\).

\subsection{Ambiguity of Vacuum}\label{ch amb va}
We should notice that our choice of mode function \(f_i\) is non-unique, and they are not associated with some symmetry group. Therefore we can always construct another set of modes of function
\(g_i\) and follow the procedure in the previous section, to construct the Fock space and vacuum state for the mode \(g_i\).


In Minkowski spacetime, the natural set of modes is associated with the Cartesian coordinate system \((t,x,y,z)\). These coordinates are associated with the Killing vectors and Poincare group.
The Fock space and vacuum state are invariant under the action of the Poincare group.
More specifically, the positive frequency mode in Minkowski spacetime is defined through the Lie derivative along the timelike Killing vector, \(\mathcal{L}_{\partial t}f_{\mathbf{k}}=-i\omega f_{\mathbf{k}}\), and this timelike Killing vector is associated with the Poincare group so that we are always able to distinguish positive frequency and negative frequency modes in
flat spacetime.

Back to the discussion in curved spacetime, let's consider another set of modes \(g_i\), we can expand the field operator in terms of \(g_i\),
\begin{eqnarray}\label{ame}
    \hat{\phi}=\sum_i\left[{}^{(2)}\hat{{a}}_i g_i+{}^{(2)}\hat{{a}}_i^\dagger g_i^*\right].
\end{eqnarray}
Follow the same procedures for mode \(f_i\), the commutation relation is imposed;
\begin{equation}
\begin{aligned}
  \bigl[{}^{(2)}\hat{{a}}_{i},{}^{(2)}\hat{{a}}_{j}\bigr]                     & =0    ,       \\
  \bigl[{}^{(2)}\hat{{a}}^{\dagger}_{i},{}^{(2)}\hat{{a}}^{\dagger}_{j}\bigr] & =0   ,        \\
  \bigl[{}^{(2)}\hat{{a}}_{i},{}^{(2)}\hat{{a}}^{\dagger}_{j}\bigr]           & =\delta_{ij}.
\end{aligned}
\end{equation}
The vacuum state of this set of annihilation operators is
\begin{eqnarray}
    \forall\;i,\ {}^{(2)}\hat{{a}}_i\ket{0_g}=0.
\end{eqnarray}
The Fock basis is constructed by repeated action of creation operators on this vacuum, and the number operator is defined as
\begin{eqnarray}
    \hat{n}_{g_i}={}^{(2)}\hat{{a}}_i^\dagger {}^{(2)}\hat{{a}}_i.
\end{eqnarray}
Both sets of modes are constructed in an equal way, and we don't know which one defines a vacuum that is closer to the physical vacuum.

Now, since both sets are complete, we can express one in terms of another, known as \textbf{Bogoliubov transformation};
\begin{equation}
\begin{aligned}\label{bt}
    f_i & =\sum_j\left[\alpha^*_{ji}g_j-\beta_{ji}g^*_j\right] ,\\
    g_i & =\sum_j\left[\alpha_{ij}f_j+\beta_{ij}f_j^*\right].
\end{aligned}
\end{equation}
The matrix element \(\alpha_{ij}\) and \(\beta_{ij}\) are called \textbf{Bogoliubov coefficients}.
Using the orthogonality of mode functions, we can compute the Bogoliubov coefficients,
\begin{equation}
\begin{aligned}
    \alpha_{ij} & =(g_i,f_j)    , \\
    \beta_{ij}  & = -(g_i,f_j^*).
\end{aligned}
\end{equation}
In order to make a physical mode function, the Bogoliubov coefficients should be normalized,
\begin{equation}
\begin{aligned}
  \sum_j\bigl[\alpha_{ik}\alpha^*_{jk}-\beta_{ik}\beta^*_{jk}\bigr] & =\delta_{ij}, \\
  \sum_j\bigl[\alpha_{ik}\beta_{jk}-\beta_{ik}\alpha_{jk}\bigr]     & =0.
\end{aligned}
\end{equation}
To obtain the relation of creation/annihilation operators of set \(f_i\) and \(g_i\),
we first equating the mode expansions Eq. \eqref{cme} and Eq. \eqref{ame}, and use Bogoliubov transformation Eq. \eqref{bt};
\begin{equation}
  \sum_i\left[{}^{(1)}\hat{a}_i f_i+{}^{(1)}\hat{a}_i^\dagger f_i^*\right]=\sum_i\left[{}^{(2)}\hat{{a}}_i g_i+{}^{(2)}\hat{{a}}_i^\dagger g_i^*\right].
\end{equation}
We have
\begin{equation}
\begin{aligned}
  {}^{(1)}\hat{a}_i       & =\sum_j\left[\alpha_{ji}{}^{(2)}\hat{{a}}_j+\beta^*_{ji}{}^{(2)}\hat{{a}}^{\dagger}_j\right] \\
  {}^{(2)}\hat{{a}}_i & =\sum_j\left[\alpha^*_{ij}{}^{(1)}\hat{a}_j-\beta^*_{ij}{}^{(1)}\hat{a}^{\dagger}_j\right].
\end{aligned}
\end{equation}
Now imagine an observer in the vacuum state defined by mode \(f_i\), we want to know how many particles are there if the state is defined by \(g_i\) mode. Then we should examine the expectation value of the number operator of \(g_i\), \(\hat{n}_{g_i}\) in \(f_i\)-vacuum,
\begin{equation}
    \begin{aligned}
          \expval{\hat{n}_{g_i}}{0_{f_i}} & =\expval{{}^{(2)}\hat{{a}}_i^\dagger {}^{(2)}\hat{{a}}_i}{0_{f_i}}                                                                                                     \\
                                  & =\expval{\sum_{jk}\left(\alpha_{ij}{}^{(1)}\hat{a}^{\dagger}_j-\beta_{ij}{}^{(1)}\hat{a}_j\right)\left(\alpha^*_{ik}\hat{a}_k-\beta^*_{ik}\hat{a}^{\dagger}_k\right)}{0_{f_i}} \\
                                  & =\sum_{jk}\left(-\beta_{ij}\right)\left(-\beta^*_{ik}\right)\expval{{}^{(1)}\hat{a}_j{}^{(1)}\hat{a}^{\dagger}_k}{0_{f_i}}                                                     \\
                                  & =\sum_{jk}\left(\beta_{ij}\beta^*_{ik}\right)\expval{{}^{(1)}\hat{a}^{\dagger}_k{}^{(1)}\hat{a}_j+\delta_{jk}}{0_{f_i}}                                                        \\
                                  & =\sum_{jk}\left(\beta_{ij}\beta^*_{ik}\right)\delta_{jk}\braket{0_{f_i}}{0_{f_i}}                                                                              \\
                                  & =\sum_{j}\beta_{ij}\beta^*_{ij}                                                                                                                                \\
                                  & =\sum_{j}|\beta_{ij}|^2.
    \end{aligned}
\end{equation}
Hence, the number of particle in mode \(g_i\) in the \(f_i\)-vacuum is equal to \(\sum_{j}|\beta_{ij}|^2\).

\subsection{Asymptotic Bounded Model of Cosmological Particle Creation}
In the early work of cosmological particle creation, researchers usually employ some specific kind of model called the asymptotic bounded model \cite{Par69,Par71,BerDun}. This kind of model includes two asymptotic static regions, in-region and out-region. In the in-region the concept of vacuum can be well-defined; in the out-region, the concept of the particle can be well-defined. Thus, we can calculate the resulting particle creation in the out-region after some time evolution between in-and-out-region. In this section we will focus on the massive conformal particle creation in the spatially flat Friedmann–Lemaître–Robertson–Walker spacetime study by Bernard and Duncan \cite{BerDun}. A more detailed review of cosmological particle production and other models can be found in \cite{Ford:2021syk}.

The metric of spatially flat Friedmann–Lemaître–Robertson–Walker spacetime can be written as
\begin{eqnarray}
    ds^2 = \dd t^2 - a^2(t)(\dd x^2+\dd y^2 +\dd z^2) = a^2(\eta)(\dd \eta^2 -\dd x^2-\dd y^2 -\dd z^2),
\end{eqnarray}
where $\dd\eta =\dd t/a(t)$ is conformal time. The definition of conformal time allows a simpler form of wave equation that will shown later. The mode function in Eq. \eqref{cme} is separable and takes the form 
\begin{eqnarray}\label{fmode}
    f_{\vb{k}}(x^{\mu}) = \frac{1}{(2\pi)^{3/2}}\frac{\chi(\eta)}{a(\eta)} e^{\mathrm{i} \vb{k}\cdot \vb{x}}.
\end{eqnarray}
Plug Eq. \eqref{fmode} back to the Klein-Golden equation Eq. \eqref{kg} with $\xi =1/6$, we have
\begin{eqnarray}\label{egchieq}
    \chi_k''+\qty[k^2 + m^2 a^2(\eta)]\chi_k = 0,
\end{eqnarray}
where prime denotes the partial derivative with respect to conformal time $\eta$. 
The normalization requirement of $f_{\vb{k}}$ become the Wronskian of Eq. \eqref{egchieq},
\begin{eqnarray}\label{wrons}
    \chi'^*_{\vb{k}}\chi^{\vphantom{*}}_{\vb{k}}-\chi_{\vb{k}}^*\chi'_{\vb{k}}=\mathrm{i}.
\end{eqnarray}
The mode function $\chi_k$ can be separated into positive and negative frequency parts,
\begin{eqnarray}
    \chi_k = \alpha_k \chi_k^{(+)} +\beta_k \chi_{k}^{(-)},
\end{eqnarray}
where $\alpha_k$ and $\beta_k$ are coefficients of positive frequency and negative frequency part respectively, and they are related to Bogoliubov coefficients by 
\begin{equation}
    \begin{aligned}
        \alpha_{\vb{k},\vb{k'}} &=\alpha_k\delta_{\vb{k},\vb{k'}}\\
        \beta_{\vb{k},\vb{k'}} &=\beta_k\delta_{\vb{k},-\vb{k'}}.
    \end{aligned}
\end{equation}
The equation \eqref{egchieq} has the form of a parametric oscillator, and an analysis of differential equations with this form reveals that the coefficients of both positive frequency and native frequency part would be amplified as time evolves. This phenomenon is named as \textbf{parametric amplification}.

\begin{figure}[htbp]
    \centering
    \includegraphics[width=0.7\textwidth]{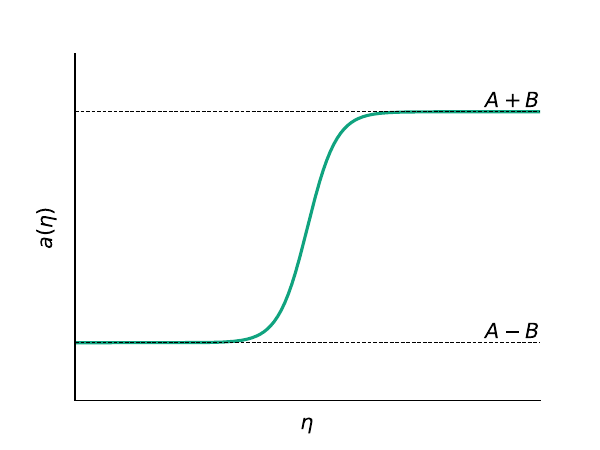}
    \caption{Scalar factor of Bernard-Duncan model. The space begins with a static period, then expands exponentially, and back to static. }
    \label{fig:BDa}
\end{figure}

The scale factor of the Bernard-Duncan model represents a spacetime with static in and out state with an exponential rise region (Fig. \ref{fig:BDa}),
\begin{eqnarray}
    a^2(\eta) = A+B\tanh(\rho\eta),
\end{eqnarray}
with $a^2(-\infty)=a_i^2=A-B$ and $a^2(\infty)=a_f^2=A+B$.
We require the quantum field at in-region to be the vacuum state, which is equivalent to requiring $\beta_k$, the coefficient of negative frequency mode, to vanish at asymptotic past. The Eq. \eqref{egchieq} can be arranged to a hypergeometric differential equation, and the positive frequency part at asymptotic past and future can be solved in terms of hypergeometric function with asymptotic boundary conditions respectively. These two sets of solutions are related by the transformation rule of the hypergeometric function, and one can obtain the both coefficients of positive frequency and negative frequency by using the Bogoliubov transformation relation \cite{BerDun}. This lead to 
\begin{eqnarray}\label{bdbeta}
    |\beta_k|^2 = \frac{\sinh^2(\pi \Omega_-/\rho)}{a_i^2 \sinh(\pi \Omega^\text{out}/\rho)\sinh(\pi \Omega_\text{in}/\rho)},
\end{eqnarray}
where 
\begin{eqnarray}
    \Omega_\pm = \frac{1}{2}(\Omega^{\text{out}}\pm \Omega_{\text{in}}),\quad \Omega^{\text{out}}=\sqrt{k^2 +m^2a_f},\quad  \Omega_{\text{in}}=\sqrt{k^2 +m^2a_i}.
\end{eqnarray}
As shown in the previous section, the expectation value of the number density operator of particles in specific frequency is given by $|\beta_k|^2$, therefore, particles are created from the vacuum by the expanding universe. 

With expression in Eq. \eqref{bdbeta}, one can further show that the particle number is proportional to $m^4$ when $m$ is small, since if $m =0$ the parametric oscillator equation Eq. \eqref{egchieq} will no longer have time-dependent frequency, the particle creation will not happen. At the limit that $m$ is large, the particle number is exponentially suppressed with $m$, which can be understood as heavier particles are harder to create.

\newpage
\section{Dynamical Casimir Effect in Two-dimension and Conformal Anomaly}\label{2d}
In this section, we work with a simple 1+1D geometry, in which the backreaction of a massless conformal quantum field is due to the Casimir energy and the trace anomaly. In the next section, we consider a 3+1 D dimensional configuration and highlight the other important kind of backreaction, due to particle creation.   

Once we assume the space between the moving mirrors is homogeneous, the size of space varying with time can be effectively described by the FLRW metric with line element
\begin{equation}
    ds^2 \equiv g_{\mu\nu} \dd x^\mu \dd x^\nu = \dd t^2 - a^2(t) \dd x^2 = a^2(\eta)(\dd \eta^2 - \dd x^2 )
    \label{2dmetric}
\end{equation} 
We fixed one mirror at coordinate position $x=0$ and allowed another mirror to move at a coordinate position $x=l$. This is equivalent to Dirichlet boundary conditions imposed at coordinate positions $x=0$ and $x=l$, and the motion of the mirror is described by scale factor $a(t)$. We consider a massive scalar field inside the cavity, whose action has the form\footnote{keep the mass here is required for the regularization of the energy-momentum tensor, and we will set the mass to $0$ once we are done with regularization.}
\begin{equation}
	S_f= \int \dd t \int_0^l \dd x \,|g|^{1/2}\frac{1}{2}\Big[g^{\mu\nu}\big(\partial_\mu\phi\big)\big(\partial_\nu\phi\big)-m^2\phi^2\Big].
\label{Eq:Sf}
\end{equation}
Here, $g \equiv \text{det}[g_{\mu\nu}] = -a^2$, while $g^{\mu\nu}$ is the inverse metric tensor: $g^{\mu\alpha}g_{\alpha\nu} = \delta_\nu^\mu$.
The overall action of this moving mirror system is 
\begin{eqnarray}\label{2dtaction}
    S = S_m + S_f = \int \dd t\;\qty[ \frac{1}{2}M\dot{a}^2 l^2] + \int \dd t \int_0^l \dd x \,|g|^{1/2}\frac{1}{2}\Big[g^{\mu\nu}\big(\partial_\mu\phi\big)\big(\partial_\nu\phi\big)-m^2\phi^2\Big],
\end{eqnarray}
where $M$ is the mass of the moving mirror and the dot denotes the partial derivative with respect to cosmic time $t$.
Variation with respect to $a(t)$ gives the equation of motion of the mirror,
\begin{equation}\label{mirrorraweom}
    M \ddot{a}l =\frac{1}{2l}\int_0^l \dd x \qty[(\partial_t\phi)^2+\frac{1}{a^2}(\partial_x\phi)^2-m^2\phi^2].
\end{equation}
Inspect the RHS of Eq. \eqref{mirrorraweom}, one can notice that it can be written in terms of the component of energy-momentum tensor (EMT),
\begin{equation}\label{2dmeom}
    M \ddot{a}l=\frac{1}{l}\int_0^l\dd x \qty[T_{tt} - T_{\rho}^{\ \rho}],
\end{equation}
where 
\begin{eqnarray}\label{2dEMT}
    T_{\mu\nu}\equiv\frac{2}{|g|}\frac{\delta S_f}{\delta g^{\mu\nu}}=(\partial_\mu\phi)(\partial_\nu\phi)-\frac{1}{2}g_{\mu\nu}(\partial^{\rho}\phi)(\partial_{\rho}\phi)+\frac{1}{2}g_{\mu\nu}m^2\phi^2.
\end{eqnarray}
Thus, the energy density and trace of the energy-momentum tensor of the field serve as the source term of the equation of motion of the mirror. 

We shall proceed to calculate the energy-momentum tensor with quantum correction. The scalar field equation is obtained by varying the action Eq. \eqref{2dtaction} with respect to $\phi$, 
\begin{eqnarray}\label{2d phieq}
    \frac{\delta S}{\delta \phi}=0&\Rightarrow& \qty[\square +m^2]\phi =0\\
    &\Rightarrow&\partial_t^2\phi +\frac{\dot{a}}{a}\partial_t\phi-\frac{1}{a^2}\partial_x^2\phi + m^2\phi =0.
\end{eqnarray}
After the canonical quantization described in the previous section, we can expand the field operator on an arbitrary basis, as
\begin{eqnarray}\label{2dexpansion}
    \hat{\phi}(x^{\mu}) = \sum_{n=1}^{\infty}\qty(\hat{a}_k f_{k_{n}}(x^{\mu})+\hat{a}_k^\dagger f_{k_{n}}^*(x^{\mu})).
\end{eqnarray}
Since we approximate the space to be homogenous, the mode function $f_k(x^{\mu})$ can be separated into spatial eigenfunction and time-dependent part,
\begin{eqnarray}\label{2dmodesep}
    f_k(x^{\mu})=\frac{1}{l^{1/2}}\chi(\eta)\sin(k_{n}x)
\end{eqnarray}
where $k_{n} = n\pi/l$. We will suppress the subscript $n$ for brevity in the following. 
Plug Eq. \eqref{2dmodesep} back to the Eq. \eqref{2d phieq} (in conformal time again) give
\begin{eqnarray}\label{2dchieq}
    \chi_k'' +(k^2 + m^2 a^2(\eta)) \chi_k =0,
\end{eqnarray}
with normalization constraint Eq. \ref{wrons}.
Noticed that the massless limit of the above equation can be solved directly with the normalization constraint. Use the relation
\begin{eqnarray}\label{EMTrela}
    \expval{T_{\mu\nu}} = \int\dd u T_{\mu\nu}[f_k,f_k^*],
\end{eqnarray}
where $\dd u$ is the measure in the mode expansion Eq. \eqref{2dexpansion} and $T_{\mu\nu}[f_k,f_k^*]$ represent the bilinear expression in Eq. \eqref{2dEMT},
we can obtain the energy density and trace of EMT, in the conformal frame,
\begin{eqnarray}\label{2dEMTo}
    \expval{T_{\eta\eta}}&=& \frac{\pi}{2l^2}\sum_{n=1}^{\infty}n\\
    \expval{T_{\rho}^{\ \rho} }&=& = 0.
\end{eqnarray}
The trace is zero for the classical conformal field as expected, and the energy density is formally divergent as we are summing over zero-point fluctuation to infinity. 

There are many ways to subtract the divergent from EMT, here we will employ a method called adiabatic regularization \cite{ParFul74,FuParHu74}, which was inspired by wave scattering in inhomogeneous media, similar to the ‘n-wave regularization’ of Zel’dovich and Starobinsky \cite{ZelStar}, and has been shown to be effective in regularizing the quantum field in a dynamical background. The procedure of adiabatic regularization can be outlined as follows;
\begin{enumerate}
    \item Keep the mass non-zero in Eq. \eqref{2dchieq}, so the frequency remain time-dependent.
    \item Take the limit $l\rightarrow \infty$, so the spectrum of mode will be continuous \cite{AndPar87}. 
    \item Use the WKB method to obtain the approximation solution of the positive frequency part of Eq. \eqref{2dchieq} up to $n$-th order in time, where $n$ is the dimension of spacetime. 
    \item Use the WKB positive frequency solution to obtain the vacuum expectation value of EMT $\expval{T_{\mu\nu}}{0_A}$.
    \item Subtract the $\expval{T_{\mu\nu}}{0_A}$ from the original expression of EMT Eq. \eqref{2dEMTo}, and take the mass to be zero. This would result in a finite physical expression of EMT, containing the quantum corrections. 
\end{enumerate}
The central idea of this method is to maximally approximate the UV part of vacuum energy density, which contains only the unphysical divergence, and leave the IR behavior unchanged. Even though, in some cases, it is possible to find the WKB approximate EMT is not divergence, it still should be subtracted, as it still belongs to the unphysical UV part of EMT. This is one of the central results of QFTCS in the 1970s, leading to the discovery of conformal anomaly.

In the limit $l\rightarrow\infty$, mode function Eq. \eqref{2dmodesep} would be
\begin{eqnarray}\label{cmodesep}
     f_k(x^{\mu})=(2\pi)^{1/2}\chi(\eta)\exp(\mathrm{i}kx)
\end{eqnarray}
with mode expansion in continuous (integral) measure
\begin{eqnarray}\label{2dcmodeexp}
    \hat{\phi}(x^{\mu}) = \int_{-\infty}^{\infty}\dd k\; \qty(\hat{a}_k f_k(x^{\mu})+\hat{a}_k^\dagger f_k^*(x^{\mu}))
\end{eqnarray}
and the same equation for $\chi_k$ Eq. \eqref{2dchieq}.

In 2-dimensional spacetime, applying the WKB iteration to Eq. \eqref{2dchieq} up to the second order should capture all the divergent terms,
\begin{equation}
    \begin{aligned}\label{2dwkb}
    \chi_k&=(2W_{k})^{-1/2}\exp{-i\int^{\eta}W_k(\eta')\dd\eta'}\\
    W_k&=\omega_k(1+\epsilon_2)^{1/2}\approx\omega_k(1+\frac{1}{2}\epsilon_2)\\
    \epsilon_2&=-\omega_k^{-1/2}\frac{\dd^2}{\dd\eta^2}\left[ \omega_k^{1/2} \right],\quad
    \omega_{k}=\sqrt{k^2+m^2a^2(\eta)}.
\end{aligned}
\end{equation}

Combining Eq. \eqref{2dwkb} and Eq. \eqref{cmodesep} and using Eq. \eqref{EMTrela}, we find an expression for WKB approximated energy density up to second-order,
\begin{equation}
    \begin{aligned}
    \expval{T_{\eta\eta}}{0_A}&=\frac{1}{4\pi}\int_{0}^{\infty} \frac{\dd k}{W_k}\left[W_k^2+k^2+\frac{1}{4}\left(\frac{W'_{k}}{W_k}\right)^2\right]\\
    &=\frac{1}{4\pi}\int_{0}^{\infty}\dd k\;\Biggl[\omega_k+\frac{k^2}{\omega_k}-\frac{\qty[2a^2(2\dot{a}^2+a\ddot{a})]m^2}{8\omega_k^3}+\frac{\qty[2a^2(2\dot{a}^2+a\ddot{a})]k^2m^2}{8\omega_k^5}\\
    &~~~~~~~~~~~~~~~~~~~~~~~~~~~~~~+\frac{7(4a^4\dot{a}^2)m^4}{32\omega^5_k}-\frac{5(4a^4\dot{a}^2)k^2m^4}{4\omega_k^7}\Biggr] \\
 &=\frac{1}{2\pi} \int_{0}^{\infty}\dd k\;\omega_k + \frac{\dot{a}^2}{24\pi }.
\end{aligned}
\end{equation}
The massless limit has been taken and the terms with order in time higher than two have been discarded. With this expression, we can perform the regularization of energy density,
\begin{equation}
    \begin{aligned}
    \expval{T_{\eta\eta}}_{\text{phys}}&=\expval{T_{\eta\eta}}-\expval{T_{\eta\eta}}{0_A}\\
    &=\frac{\pi}{2l^2}\sum_{n=0}^{\infty}n-\frac{1}{2\pi} \int_{0}^{\infty}\dd k\;\omega_k - \frac{\dot{a}^2}{24\pi }\\
    &=\lim_{\alpha\rightarrow 0}\qty[\frac{\pi}{2l^2}\sum_{n=0}^{\infty}ne^{-\alpha \pi n/l}-\frac{1}{2\pi } \int_{0}^{\infty}\dd k\;k e^{-\alpha k}] - \frac{\dot{a}^2}{24\pi }\\
    &=\lim_{\alpha\rightarrow 0}\qty[\frac{\pi}{2l^2}\frac{e^{\alpha \pi/l}}{\qty(e^{\alpha\pi/l}-1)^2}-\frac{1}{2\pi \alpha^2}]- \frac{\dot{a}^2}{24\pi}\\
    &=\lim_{\alpha\rightarrow 0}\qty[\qty(\frac{1}{2\pi \alpha^2}-\frac{\pi}{24l^2}+\frac{\pi^3\alpha^2}{480l^2}+O(\alpha^n))-\frac{1}{2\pi \alpha^2}]- \frac{\dot{a}^2}{24\pi}\\
    &=-\frac{\pi }{24 l^2} - \frac{\dot{a}^2}{24\pi }.
\end{aligned}
\end{equation}

In the third line of the above equation, we introduced a cutoff function to suppress the diverging series and integral, this method has been largely used in calculating the Casimir effect and is sometimes referred to as Heat-kernel regularization \cite{Schwartz:2014sze}. 
Transform the $\expval{T_{\eta\eta}}_{\text{phys}}$ back to the comoving frame (or lab frame),
\begin{eqnarray}
    \expval{T_{tt}}_{\text{phys}} = \frac{\partial \eta}{\partial t}\frac{\partial \eta}{\partial t}\expval{T_{\eta\eta}}_{\text{phys}} = -\frac{\pi }{24 a^2l^2} - \frac{\dot{a}^2}{24\pi a^2}.
\end{eqnarray}
So far we have obtained one term in the mirror's equation of motion Eq. \eqref{2dmeom}, next we shall calculate the regularized trace of EMT, and show that the classical conformal invariant field does not remain conformal invariant after quantization. 

Again using the WKB mode function in Eq. \eqref{2dwkb}, we can get the expectation value of the trace at the second-adiabatic order. By subtracting this from the classical value of the trace $ \expval{T_{\rho}^{\ \rho} } = 0$, we obtain
\begin{equation}
\begin{split}
	\expval{T_\rho^{\ \rho}}_{\rm phys} &= 0- m^2 \big<\phi^2\big>_{A} \\
	&=  \frac{m^2}{4\pi a}\int_{-\infty}^{+\infty} \dd k\,\frac{W_k}{\omega_k^4}\\
	&= \frac{m^2}{16\pi a}\int_{-\infty}^{+\infty} \dd k\,\frac{1}{\omega_k^3}\left(\frac{3}{2}\frac{\dot{\omega}_k^2}{\omega_k^2}-\frac{\ddot{\omega}_k}{\omega_k}-\frac{1}{2} \left(\frac{\ddot{a}}{a} - \frac{\dot{a}^2}{2a^2}\right)\right). \label{Eq:TAn}
\end{split}
\end{equation}
Taking the massless limit $m\to 0$ of this expression, and by noting that the integral in Eq. \eqref{Eq:TAn} does not depend on the value of $m$ [this can be seen by changing the variable of integral to $k/(ma)$], the anomaly trace takes the simple form,
\begin{equation}
    \expval{T_\rho^{\ \rho}}_{\rm phys} = -\frac{1}{12\pi}\frac{\ddot{a}}{a}.
\end{equation}
This result confirms that the physical value of the trace is equal to $\expval{T_\rho^{\ \rho}}_{\rm phy}=R/(24\pi)$, with $R$ the scalar curvature of metric Eq. \eqref{2dmetric}, a coordinate invariant quantity, as expected.

So far we have everything to build the mirror's equation of motion,
\begin{eqnarray}
       M \ddot{a}l&=&\frac{1}{l}\int_0^l\dd x \qty[T_{tt} - T_{\rho}^{\ \rho}] \\
       &=&\frac{1}{l}\int_0^l\dd x \qty[-\frac{\pi }{24 a^2l^2} - \frac{\dot{a}^2}{24\pi a^2} +\frac{1}{12\pi}\frac{\ddot{a}}{a}]\\
       &=& -\frac{\pi }{24 a^2l^2} - \frac{\dot{a}^2}{24\pi a^2} +\frac{1}{12\pi}\frac{\ddot{a}}{a}.
\end{eqnarray}
Recover the physics length of cavity $L = al$, we have
\begin{eqnarray}\label{2demof}
    M \ddot{L}= -\frac{\pi }{24 L^2} - \frac{\dot{L}^2}{24\pi L^2} +\frac{1}{12\pi}\frac{\ddot{L}}{L}.
\end{eqnarray}
This result agrees with the general form of backreaction force of DCE in one dimension obtained by many authors, e.g. \cite{DODONOV1989511,Beyer:1990dq},
\begin{eqnarray}
    M \ddot{L}= -\frac{\pi }{24 L^2} - A\frac{\dot{L}^2}{L^2} +B\frac{\ddot{L}}{L}.
\end{eqnarray}
The numerical coefficients $A$ and $B$ are different by those authors due to different situations and different approximations they employed. This shows our homogeneous approximation correctly captured the qualitative behavior of the backreaction of DCE. 

\begin{figure}[htbp]
    \centering
    \includegraphics[width=0.7\textwidth]{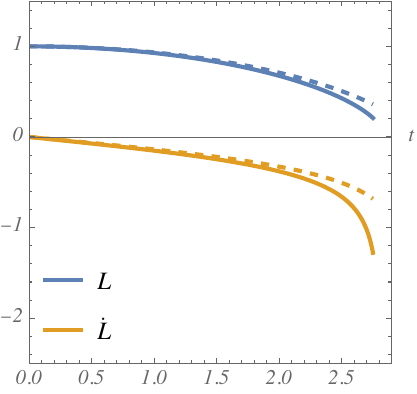}
    \caption{Plot shows the backreaction of DCE in a one-dimensional cavity is in the same direction of static Casimir force, speeding up the collapse of two mirrors. The mass of the mirror is set to $1$.}
    \label{fig:2dDce}
\end{figure}

The Eq. \eqref{2demof} is numerically solved, shown in Fig. \ref{fig:2dDce}. To compare, we also plot the motion of the mirror without the DCE modification, i.e., the mirror moving only under static Casimir force, shown as dashed lines. 

\newpage
\section{Dynamical Casimir Effect in Four-dimension: Particle Production}\label{4d}
In this section, we will deal with a more realistic model, the dynamical Casimir effect in a perfect conduction box with one side moving. Again we approximate the quantum field inside the box to be spatially homogeneous. 

This time we can approximate the space inside the cavity by a simplified version of the Bianchi-type I metric,
\begin{equation}\label{3dmetric}
    ds^2 \equiv g_{\mu\nu} dx^\mu dx^\nu= \dd t^2 - \qty[a_1^2(t)\dd x^2 + a_2^2(t)\dd y^2 + a_3^2(t)\dd z^2],
\end{equation}
where $a_i$ are the scale factors in the three directions. We consider the case with $a_2=a_3=1$, $a_1=a(t)$, i.e., with only one side moving, in the $\pm x$ direction.  Imposing Dirichlet boundary conditions at both ends of each of the three sides, i.e., between $x,y,z = 0$ and $x,y,z = l$, where $l$ is the length would turn this into a `box universe'. Our present setup has $l$ fixed in the $y, z$ directions, but the scale factor in the $x$ direction is allowed to change in time. 

Consider a massless conformal scalar field $\phi$ (the corresponding massless conformal vector field would describe a photon field) in this box `universe'  with action
\begin{align}
    S_f = \frac{1}{2}\int dx^4 \sqrt{|g|} \qty[g^{\mu\nu}\partial_{\mu}\phi\partial_{\nu}\phi -\frac{1}{6}R\phi^2]
\end{align}
obeying the Klein-Gordon equation,
\begin{align}
    &~~~~~\square\phi +\frac{1}{6}R\phi = 0,\\
    &\Rightarrow \partial_t^2\phi + \frac{\dot{a}}{a}\dot{\phi} - \sum_i\frac{\partial_i^2\phi}{a_i^2}+\frac{1}{6}R\phi=0, 
\end{align}
where $R$ is the scalar curvature of metric Eq. \eqref{3dmetric}.

We will take a slightly different approach from the previous chapter to obtain the equation of motion of the mirror, and the reason will be clear later. We will first calculate the regularized energy density from the energy-momentum tensor, and use this energy density to build an ``effective action'' to obtain the equation of motion. The energy-momentum tensor of this scalar field is given by
\begin{align}
    T_{\mu\nu} &\equiv{ \frac{2}{\sqrt{|g|}}\frac{\delta S}{\delta g^{\mu\nu}}}= (\partial_\mu\phi)(\partial_\nu\phi)-\frac{1}{2}g_{\mu\nu}g^{\lambda\sigma}(\partial_\lambda\phi)(\partial_\sigma\phi) \nonumber\\
    &-\frac{1}{6}\left[\nabla_\mu\partial_\nu(\phi^2)-g_{\mu\nu}g^{\lambda\sigma}\nabla_\lambda\partial_\sigma(\phi^2)+\phi^2G_{\mu\nu}\right],
\end{align} 
where $G_{\mu\nu}\equiv R_{\mu\nu}-(1/2)g_{\mu\nu}R$ is the Einstein tensor. We are interested in the  energy density 
\begin{equation}
    T_{tt}=\frac{1}{2}(\dot{\phi})^2 + \frac{1}{3}\frac{\dot{a}}{a}\phi\dot{\phi}+\frac{1}{6}\sum_i \left(\frac{\partial_i\phi}{a_i}\right)^2-\frac{1}{3}\sum_i \left(\frac{\phi\partial_i^2\phi}{a_i^2}\right)-\frac{1}{6}G_{tt}\phi^2,
\end{equation}
where an overdot indicates differentiation with respect to $t$.
Introduce a new field variable $\chi$ and a conformal time variable $\eta$,
\begin{equation}
    \chi = a^{1/3}\phi, \; \dd\eta =  a^{-1/3}{\dd t}.
\end{equation}
In these variables, one gets a conformally-flat metric when the spacetime is isotropic. In our case, it can eliminate the first-order term in the differential equation of the time-dependent part of the mode function. 
By substituting 
\begin{equation}
    \dot{\phi} = a^{-2/3}\left[ \chi' - \frac{1}{3}\frac{a'}{a}\chi\right]
\end{equation} 
(where a prime indicates differentiation with respect to $\eta$) into the above equation for the energy density, we have
\begin{equation}
    T_{00}= a^{-2/3}\qty[\frac{1}{2}a^{-2/3}(\chi')^2 +\frac{1}{6}\sum_i\qty(\frac{\partial_i\chi}{a_i})^2-\frac{1}{3}\sum_i\left(\frac{\chi\partial_i^2\chi}{a_i^2}\right)-\frac{1}{2}a^{-2/3}Q\chi^2],
\end{equation}
where  $Q$ is defined as 
\begin{equation}
    Q = \frac{1}{18}\sum_{i<j}\qty(\frac{a_i'}{a_i}-\frac{a_j'}{a_j})^2= \frac{1}{9}\qty(\frac{a'}{a})^2
\end{equation}
representing the degree of anisotropy. Now the scalar field equation becomes 
\begin{equation}
    \chi'' - a^{2/3}\sum_i \frac{2\partial_i^2\chi}{a_i^2}+Q\chi = 0
\end{equation}
Although this spacetime is anisotropic, it remains homogeneous. Under the Dirichlet boundary condition imposed at the boundaries, we can perform the following mode decomposition,
\begin{equation}
    \chi = \frac{1}{l^{3/2}}\sum_\mathbf{k} \qty[A_{\vec{k}}\chi_{\vec{k}}(\eta)w_{\vec{k}}(x^i)+A_{\vec{k}}^{\dagger}\chi_{\vec{k}}^*(\eta)w_{\vec{k}}(x^i)]
\end{equation}
with the compact notation $$\sum_{\vb{k}}= \prod_{i=1}^{3}\sum_{n_i}$$ and
\begin{eqnarray}
    w_{\vec{k}}(x^i) = \sin(\frac{n_x \pi x}{l})\sin(\frac{n_y \pi y}{l})\sin(\frac{n_z \pi z}{l})
\end{eqnarray}
with $n_i = 1, 2, 3, \cdots$.
The Fourier modes $\chi_{\Vec{k}}$ satisfy the parametric oscillator equation:
\begin{align}
    \chi_{\Vec{k}}''+(\Omega_{\Vec{k}}^2+Q)\chi_{\vec{k}}=0
    \label{xeq}
\end{align}
with time-dependent frequency 
\begin{eqnarray}
        \Omega_{\Vec{k}}^2=\qty(a^{1/3}\omega_{\vec{k}})^2&=&a^{2/3}\qty(\sum_i\frac{k_i^2}{a_i^2}+m^2)=a^{2/3}\qty(\frac{k_x^2}{a^2}+k_y^2+k_z^2+m^2),\\
        k_i&=&\frac{n_i \pi x^i}{l}.
\end{eqnarray}

Let $\ket{0_A}$ be an arbitrary vacuum state defined by $A_{\vec{k}}$, which will be specified by the boundary condition of Eq. \eqref{xeq}. The vacuum expectation value of the $tt$ component of the stress-energy tensor 
\begin{equation}
    \expval{T_{tt}}{0_A}= {\frac{1}{2l^3a^{4/3}}}\sum_{\vb{k}}\bqty{\vqty{\chi_{\vec{k}}'}^2+(\Omega_{\vec{k}}^2-Q)\vqty{\chi_{\vec{k}}}^2}
    \label{div}
\end{equation}
gives the sought energy density. However, the above expression of energy density diverges due to high frequency (UV) contributions. One can remove the divergences by adiabatic regularization procedures described in \cite{FulParHu74} which we adopt in what follows (subscript $k$ on $\Omega$ are dropped). \footnote{Due to the imposition of boundary conditions the global topology of our setup is not the same as the Bianchi type-I Universe. However, UV divergence is a local property not affected by the global topology of spacetime.}
\begin{align}\label{divt}
    \rho_{\rm div} \equiv \expval{T_{tt}}_{\text{div}}&=\frac{1}{32\pi^3a^{4/3}}\int d^3k\,\Omega^{-1}\biggl\{\biggl.2\Omega^2+\qty[\frac{1}{4}\qty(\frac{\Omega'}{\Omega})^2-Q]\nonumber\\
    ~~~~~&-\frac{1}{8}\qty(\frac{\Omega'}{\Omega})^2\epsilon_{2(2)}+\frac{1}{4}\frac{\Omega'}{\Omega}\epsilon'_{2(3)}+\frac{1}{4}\Omega^2\epsilon_{2(2)}^2+\frac{1}{2}Q\epsilon_{2(2)}\biggl\},
\end{align}
where the $\epsilon_{n}$ functions are called adiabatic frequency corrections -- the subscript $n$ indicates the adiabatic order (in this case, second order) and the number in the parentheses indicates the order of time derivatives. With the definition $X = \Omega^2$ they are given by:
\begin{align}
     { \epsilon_{2(2)} }&{ =-\frac{1}{4}X^{-2}X'' + \frac{5}{16}X^{-3}(X')^2 + X^{-1}Q, }\\
    {\epsilon'_{2(3)}}&{=-\frac{1}{4}X^{-2}X''' + \frac{9}{8}X^{-3}X''X' -\frac{15}{16}X^{-4}(X')^3 + X^{-1}Q' -X^{-2}X'Q,} 
\end{align}
The regularized energy density from particle creation is obtained by subtracting Eq. \eqref{divt} from Eq. \eqref{div}
\begin{multline}
    \rho_{\rm reg} \equiv \expval{T_{tt}}_{\rm reg}=\frac{1}{2l^3 a^{4/3}}{\sum_{\vb{k}}}\bqty{\vqty{\chi_{\vec{k}}'}^2+(\Omega^2-Q)\vqty{\chi_{\vec{k}}}^2}\\
    -\frac{1}{32\pi^3a^{4/3}}\int d^3k\,\Omega^{-1}\Biggl\{\biggl.2\Omega^2 +\qty[\frac{1}{4}\qty(\frac{\Omega'}{\Omega})^2-Q]\\
   -\frac{1}{8}\qty(\frac{\Omega'}{\Omega})^2\epsilon_{2(2)}+\frac{1}{4}\frac{\Omega'}{\Omega}\epsilon'_{2(3)}+\frac{1}{4}\Omega^2\epsilon_{2(2)}^2+\frac{1}{2}Q\epsilon_{2(2)}\Biggl\}.
    \label{fullEnergy}
\end{multline}
This expression looks too complicated and involves the integral and summation that can't be evaluated analytically, which would prevent solving the equation of motion of the mirror, even numerically. Even further, $\chi_k$ in the first line requires a given mirror trajectory to be solved, which is not available before $\chi_k$ gets solved! This sounds like a deadlock. 

We will first simplify the expression, and then use an ``early time approximation'' to find a valid approximate solution of $\chi_k$ under certain constraints. Note, although Eq. \eqref{xeq} seems to allow the use of the WKB method to find an approximation solution, the WKB method does not generally give a good approximation of the particle production problem since the particle production is a ``non-adiabatic'' effect.

Inspect the Eq. \eqref{xeq}, we could require the solution of $\vqty{\chi_{\vb{k}}}$ and $\vqty{\chi_{\vb{k}}'}$ take a specific form,
\begin{align}
    \chi_{\vb{k}}&=\frac{1}{\sqrt{2\Omega_{\vb{k}}(\eta)}}\,\qty[\alpha_{\vb{k}}^{\vphantom{+}} e_{\vb{k}}^{(-)}+\beta_{\vb{k}}^{\vphantom{+}} e_{\vb{k}}^{(+)}]\,,
    &\chi'_{\vb{k}}&=-i\,\sqrt{\frac{\Omega_{\vb{k}}(\eta)}{2}}\,\qty[\alpha_{\vb{k}}^{\vphantom{+}} e_{\vb{k}}^{(-)}-\beta_{\vb{k}}^{\vphantom{+}} e_{\vb{k}}^{(+)}],
    \label{bianchi1d}
\end{align}
for two new time-dependent functions $\alpha_{k}(t)$ and $\beta_{k}(t)$, with 
\begin{equation}
    e_{\vb{k}}^{(\pm)} \equiv \exp\qty[\pm i \int_{\eta_0}^{\eta}\!\dd \eta'\;\Omega_{\vb{k}}(\eta') ]\,.
\end{equation}
The normalization condition requires $\qty|\alpha_k(\eta)|^2 - \qty|\beta_k(\eta)|^2=1$. Combining Eqs. \eqref{xeq} and \eqref{bianchi1d}, we obtain a simultaneous set of first-order complex-valued differential equations for $\alpha_{\vb{k}}$ and $\beta_{\vb{k}}$,

\begin{align}
    \alpha'_{\vb{k}}&=\frac{1}{2}\qty(\frac{\Omega'_{\vb{k}}}{\Omega_{\vb{k}}}-i\,\frac{Q}{\Omega_{\vb{k}}}) (e_{\vb{k}}^{+})^2\,\beta_{\vb{k}}-i\frac{Q}{2\Omega_{\vb{k}}}\,\alpha_{\vb{k}}\,,\label{4dal}\\
    \beta'_{\vb{k}}&=\frac{1}{2}\qty(
\frac{\Omega'_{\vb{k}}}{\Omega_{\vb{k}}}+i\,\frac{Q}{\Omega_{\vb{k}}}) (e_{\vb{k}}^{-})^2\,\alpha_{\vb{k}}+i\,\frac{Q}{2\Omega_{\vb{k}}}\,\beta_{\vb{k}}\,.\label{4dbe}
\end{align}
Now we introduce three real variables
\begin{align}
    s_{\vb{k}} &= |\beta_{\vb{k}}|^2\,,&p_{\vb{k}} &= \alpha_{\vb{k}}^{\vphantom{*}}\beta_{\vb{k}}^* e_{-}^2+\alpha_{\vb{k}}^*\beta_{\vb{k}}^{\vphantom{*}} e_{+}^2\,,&q_{\vb{k}} &=i\,(\alpha_{\vb{k}}^{\vphantom{*}}\beta_{\vb{k}}^* e_{-}^2 - \alpha_{\vb{k}}^*\beta_{\vb{k}}^{\vphantom{*}} e_{+}^2)\,,
\end{align}
thereby obtaining a simultaneous set of three first-order real-valued ordinary differential equations (four real-valued equations for two complex-valued equations, minus one from the Wronskian/normalization condition)
\begin{align}
    s'_{\vb{k}} &= \frac{1}{2}\frac{\Omega'_{\vb{k}}}{\Omega_{\vb{k}}}p_{\vb{k}} + \frac{1}{2}\frac{Q}{\Omega_{\vb{k}}}q_{\vb{k}}\,,\\
    p'_{\vb{k}} &= \frac{\Omega'_{\vb{k}}}{\Omega_{\vb{k}}}(1+2s_{\vb{k}})-\qty(\frac{Q}{\Omega_{\vb{k}}}+2\Omega_{\vb{k}})q_{\vb{k}}\,,\\
    q'_{\vb{k}} &= \frac{Q}{\Omega_{\vb{k}}}(1+2s_{\vb{k}})+\qty(\frac{Q}{\Omega_{\vb{k}}}+2\Omega_{\vb{k}})p_{\vb{k}}\,.
\end{align}
Expressing the $\chi_k$ and $\chi'_k$ in terms of $s_{\vb{k}}$, $p_{\vb{k}}$, and $q_{\vb{k}}$ and grouping the corresponding divergent and regularization terms, we find
\begin{multline}\label{E:bgivta}
    \rho_{\rm reg}={\frac{a^{-\frac{4}{3}}}{2l^3}}\sum_{\vb{k}}2s_{\vb{k}}\Omega_{\vb{k}} + \qty{{\frac{a^{-\frac{4}{3}}}{2l^3}}\sum_{\vb{k}}\Omega_{\vb{k}} -\frac{a^{-\frac{4}{3}}}{2}\int\!\frac{d^3 k}{(2\pi)^3}\;\Omega_{\vb{k}}}\\
    \qquad+\qty{{\frac{a^{-\frac{4}{3}}}{2l^3}}\sum_{\vb{k}}\qty[-\frac{Q}{2\Omega_{\vb{k}}}] -\frac{a^{-\frac{4}{3}}}{2}\int\!\frac{d^3 k}{(2\pi)^3}\;\qty[-\frac{Q}{2\Omega_{\vb{k}}}]}\\
    \qquad+\qty{{\frac{a^{-\frac{4}{3}}}{2l^3}}\sum_{\vb{k}}\qty[-\frac{Q}{2\Omega_{\vb{k}}}(2s_{\vb{k}}+p_{\vb{k}})]}
    -{\frac{a^{-\frac{4}{3}}}{2}\int\!\frac{d^3 k}{(2\pi)^3}\;\qty[\frac{1}{8}\frac{\Omega'{}_{\vb{k}}^2}{\Omega_{\vb{k}}^3}]}-\rho_{(4)}\,.
\end{multline}
where 
\begin{equation}
    \begin{aligned}
    \rho_{(4)}&=a^{-\frac{4}{3}}\int\!\frac{d^3k}{(2\pi)^3}\; \qty[ \frac{\Omega'^2_{\vb{k}}}{16\Omega_{\vb{k}}^3}-\frac{Q}{4\Omega^{\vphantom{2}}_{\vb{k}}}\qty(\frac{\Omega'^2_{\vb{k}}}{8\Omega_{\vb{k}}^4}+\frac{\Omega''_{\vb{k}}}{4\Omega^{3}_{\vb{k}}}-\frac{\Omega'_{\vb{k}}}{2\Omega_{\vb{k}}^4}-\frac{Q}{2\Omega_{\vb{k}}^2})+\Omega_{\vb{k}}^{\vphantom{2}} s_{\vb{k}}^{(4)}]\,,\\
       s_{\vb{k}}^{(4)} &= \frac{1}{16}\Biggl\{\Biggr.\frac{Q^2}{\Omega_{\vb{k}}^4}+\frac{Q'\Omega'_{\vb{k}}}{\Omega^5_{\vb{k}}}-\frac{Q}{\Omega^3_{\vb{k}}}\qty(\frac{\Omega'_{\vb{k}}}{\Omega^2_{\vb{k}}})'-\frac{3Q}{\Omega^2_{\vb{k}}}\qty(\frac{\Omega'_{\vb{k}}}{\Omega^2_{\vb{k}}})^2-\frac{1}{2\Omega^2_{\vb{k}}}\qty(\frac{\Omega'_{\vb{k}}}{\Omega^2_{\vb{k}}})''\frac{\Omega'_{\vb{k}}}{\Omega^2_{\vb{k}}}\\
    &\qquad\qquad\qquad+\frac{1}{4\Omega^2_{\vb{k}}}\qty[\qty(\frac{\Omega'_{\vb{k}}}{\Omega^2_{\vb{k}}})']^2+\frac{1}{2\Omega^{\vphantom{2}}_{\vb{k}}}\qty(\frac{\Omega'_{\vb{k}}}{\Omega^2_{\vb{k}}})'\qty(\frac{\Omega'_{\vb{k}}}{\Omega^2_{\vb{k}}})^2+\frac{3}{16}\qty(\frac{\Omega'_{\vb{k}}}{\Omega^2_{\vb{k}}})^4\Biggl.\Biggr\}\,.
\end{aligned}
\end{equation}

The first term in the first line of Eq. \eqref{E:bgivta} on the right-hand side is the energy density of the created particles, as shown in Chapter \ref{ch amb va}; the two terms in the curly bracket are the Casimir energy density due to the compactness of the spatial section.

For the terms in the second line of \eqref{E:bgivta}, we are subtracting a triple sum
\begin{align}
    -\frac{a^{-\frac{4}{3}}}{2l^3}\sum_{\vb{k}}\frac{Q}{2\Omega_{\vb{k}}}
    =-\frac{a^{-\frac{4}{3}}}{4l^3}\sum_{n_x=1}^{\infty}\sum_{n_y=1}^{\infty}\sum_{n_z=1}^{\infty}\frac{Q}{\sqrt{a^{\frac{2}{3}}\qty[\frac{(2\pi n_x)^2}{l^2}+\frac{(2\pi n_y)^2}{l^2}+\frac{(2\pi n_z)^2}{l^2}]}}\,, \label{shear energy}
\end{align}
by a triple-iterated integral. This can be done by an iterative application of the Abel-Plana summation method \cite{BordagBook},
\begin{equation}
    \sum_{n=0}^\infty f(n)=\frac 1 2 f(0)+ \int_0^\infty f(x) \, dx+ i \int_0^\infty \frac{f(i t)-f(-i t)}{e^{2\pi t}-1}\,,
\end{equation}
After regularization, the terms in the second line of \eqref{E:bgivta} in a physically reasonable case (length scale much higher than the Planck scale) give very small values, compared to the other contributions in the vacuum energy density, so we neglect these terms.

The Casimir energy in various configurations has been summarized in e.g.~\cite{BordagBook}.
The Casimir density in a rectangular box is given by
\begin{equation}\label{4dcas}
    \rho_{\textsc{ca}}(\ell_1,\ell_2,\ell_3) = -\frac{\pi^2\ell_2 \ell_3}{1440\ell_1^3}-\frac{\pi}{96\ell_1}+\frac{\zeta_{R}(3)(\ell_2+ \ell_3)}{32\pi \ell_1^2}-\frac{\pi}{2\ell_1}\qty[G\qty(\frac{\ell_2}{\ell_1})+G\qty(\frac{\ell_3}{\ell_1})]-\frac{1}{\ell_1}R\qty(\frac{\ell_2}{\ell_1},\frac{\ell_3}{\ell_1}),
\end{equation}
where $\ell_1$, $\ell_2$, $\ell_3$ are the side lengths of the rectangular box, the modified Bessel function of the second kind is given by
\begin{align}
    K_{\nu}(z) &= \frac{(z/2)^\nu \Gamma(1/2)}{\Gamma(\nu + 1/2)}\int_1^\infty\!dt\; e^{-zt} (t^2 -1 )^{(2\nu-1)/2}\,, \\
    \intertext{and}
    G(z) &= -\frac{1}{2\pi}\sum_{n=1}^{\infty}\sum_{l=1}^{\infty}\frac{n}{l}\,K_1(2\pi n l z)\,,\\
    R(z_1, z_2) &= \frac{z_1 z_2}{8}\sum_{l,p=-\infty}^{\infty}(1-\delta_{l0}\delta_{p0})\sum_{j=1}^{\infty}\qty(\frac{j}{\sqrt{l^2 z_1^2+p^2 z_2^2}})^{3/2}K_{3/2}\qty(2\pi j \sqrt{l^2 z_1^2 + p^2 z_2^2})\,.
\end{align}

\begin{figure}[htbp]
    \centering
    \includegraphics[width=0.7\textwidth]{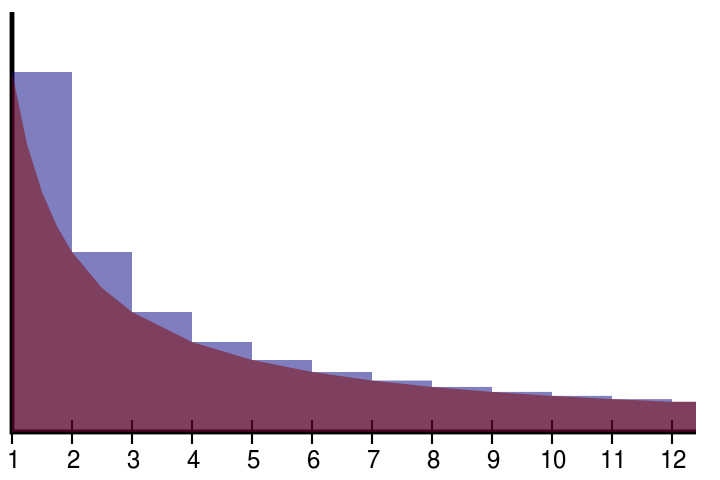}
    \caption{The illustration of Euler–Mascheroni constant from Wikipedia. This constant is the difference between harmonic series and $\int \dd x/x$, shows that the most contribution is from the ``IR'' region.}
    \label{fig:gamma}
\end{figure}

Now, energy density can be written as 
\begin{equation}\label{appro ed}
    \begin{aligned}
            \expval{\hat{T}_{tt\;\text{reg}}}{0_A}={\frac{1}{a^{\frac{4}{3}}l^3}}\sum_{\vb{k}}\bqty{s_{\vb{k}}\Omega_{\vb{k}}-\frac{Q}{4\Omega_{\vb{k}}}(2s_{\vb{k}}+p_{\vb{k}})} +\rho_{\textsc{CA}}~~~~~~~~~~~~~~~~~~~~~~\\
    \qquad-\frac{1}{a^{\frac{4}{3}}}\int\!\frac{d^3k}{(2\pi)^3}\; \Biggl[\Biggr. \frac{\Omega'^2_{\vb{k}}}{16\Omega_{\vb{k}}^3}-\frac{Q}{4\Omega^{\vphantom{2}}_{\vb{k}}}\qty(\frac{\Omega'^2_{\vb{k}}}{8\Omega_{\vb{k}}^4}+\frac{\Omega''_{\vb{k}}}{4\Omega^{3}_{\vb{k}}}-\frac{\Omega'_{\vb{k}}}{2\Omega_{\vb{k}}^4}-\frac{Q}{2\Omega_{\vb{k}}^2}) \\
    +\frac{\Omega_{\vb{k}}^{\vphantom{2}}}{16}\Biggl\{\Biggr.\frac{Q^2}{\Omega_{\vb{k}}^4}+\frac{Q'\Omega'_{\vb{k}}}{\Omega^5_{\vb{k}}}-\frac{Q}{\Omega^3_{\vb{k}}}\qty(\frac{\Omega'_{\vb{k}}}{\Omega^2_{\vb{k}}})'-\frac{3Q}{\Omega^2_{\vb{k}}}\qty(\frac{\Omega'_{\vb{k}}}{\Omega^2_{\vb{k}}})^2-\frac{1}{2\Omega^2_{\vb{k}}}\qty(\frac{\Omega'_{\vb{k}}}{\Omega^2_{\vb{k}}})''\frac{\Omega'_{\vb{k}}}{\Omega^2_{\vb{k}}}\\
    +\frac{1}{4\Omega^2_{\vb{k}}}\qty[\qty(\frac{\Omega'_{\vb{k}}}{\Omega^2_{\vb{k}}})']^2+\frac{1}{2\Omega^{\vphantom{2}}_{\vb{k}}}\qty(\frac{\Omega'_{\vb{k}}}{\Omega^2_{\vb{k}}})'\qty(\frac{\Omega'_{\vb{k}}}{\Omega^2_{\vb{k}}})^2+\frac{3}{16}\qty(\frac{\Omega'_{\vb{k}}}{\Omega^2_{\vb{k}}})^4\Biggl.\Biggr\}
    \Biggl.\Biggr]\,.
    \end{aligned}
\end{equation}
The first term in the integral diverges linearly, and it is supposed to regularize the hidden linear divergent structure in the first term of summation. The rest terms in the integral diverge logarithmically, and it is supposed to regularize the hidden logarithmic divergent structure in the second term of summation. 
However, the contribution to the difference between linearly divergent series and integrals vanishes at the asymptotic region, and this is also true for the logarithmic case. In Fig. \ref{fig:gamma}, we give a simple illustration of how this works. 

Therefore, by the reasoning we presented here, we only need to evaluate the series and integral in the low-frequency region. Evaluating the integral in Eq. \eqref{appro ed} for $k_i < 20$ gives a very small value in physically reasonable cases, allowing us to neglect this regularization contribution. 

At this step, our energy density has been simplified tremendously. What we are left with is 
\begin{eqnarray}
    \expval{\hat{T}_{tt\;\text{reg}}}{0_A}={\frac{1}{a^{\frac{4}{3}}l^3}}\sum_{\vb{k}<\Lambda}\bqty{s_{\vb{k}}\Omega_{\vb{k}}-\frac{Q}{4\Omega_{\vb{k}}}(2s_{\vb{k}}+p_{\vb{k}})} + \rho_{\textsc{CA}},
\end{eqnarray}
which is equivalent to 
\begin{eqnarray}\label{4d density1}
    \frac{1}{16\pi^3a^{4/3}}\int_{\vb{k}<\Lambda} d^3k\bqty{\vqty{\chi_{\vec{k}}'}^2+(\Omega^2-Q)\vqty{\chi_{\vec{k}}}^2-\Omega}+ \rho_{\textsc{CA}}
\end{eqnarray}
up to some negligible quantities. The continuous limit is taken, allowing us to choose a smooth low-frequency range.  

Back to the differential equation of $\chi_k$, the Eq. \eqref{4dal} and \eqref{4dbe} can be solved under the low-frequency and early time approximation $\omega_{\mathbf{k}}t < 1$. The integral like $\int_{\eta_0}^{\eta}\Omega\dd\eta'$ can be approximated to $0$. Thereby we can find the general solutions of $\alpha$ and $\beta$:
\begin{align}
    \alpha&=c_1\qty(\Omega^{1/2}-i\Omega^{-1/2}\int^{\eta}_{\eta_0}Q\dd\eta')+c_2\Omega^{-1/2},\\
    \beta&=c_1\qty(\Omega^{1/2}+i\Omega^{-1/2}\int^{\eta}_{\eta_0}Q\dd\eta')-c_2\Omega^{-1/2},
\end{align}
where $c_1(k)$ and $c_2(k)$ are complex numbers satisfying 
\begin{equation}
    c_1c_2^* + c_2c_1^* = \frac{1}{2}
\end{equation}
under the normalization condition. With this, the integrand in the first integral in Eq.~\eqref{4d density1} can be solved for $\chi_k$,
\begin{align}
    \vqty{\chi_{\vec{k}}'}^2+(\Omega^2&-Q)\vqty{\chi_{\vec{k}}}^2=2|c_1|^2\qty[(\Omega^2-Q)+\qty(\int_{\eta_0}^{\eta}Q\dd\eta')^2]\nonumber\\
    &+2|c_2|^2+2i(c_1^*c_2-c_1c_2^*)\qty(\int_{\eta_0}^{\eta}Q\dd\eta').
\end{align}
Regarding the initial conditions,  by choosing the state that corresponds to the absence of particles at time $t_0$, we can impose the initial condition
\begin{equation}
    \alpha_k(t_0) = 1,\;\beta_k(t_0) = 0.
\end{equation}
This determine the vacuum state at $t_0$, and the values of $c_1$ and $c_2$ are uniquely determined,
\begin{equation}
    c_1 =\frac{1}{2}\Omega_0^{-1/2}, \; c_2 = \frac{1}{2}\Omega_0^{1/2},
\end{equation}
where $\Omega_0$ is the value of $\Omega_k$ at $t_0$. 
Then we have
\begin{align}
    \rho&=\frac{1}{8\pi^3a^{4/3}}\int\displaylimits_{\mathcal{R}(t)} d^3k\Biggl\{ \Biggr. \qty[\frac{(\Omega^2-Q)}{\Omega_0}+\frac{\qty(\int_{\eta_0}^{\eta}Q\dd\eta')^2}{\Omega_0}]
    +\Omega_0-2\Omega\Biggl.\Biggr\} + \rho_{\textsc{CA}}.
\end{align}
Examining the term containing $Q$ integrated over time, the range of integration is suppressed if we only consider the early time behavior of the system, and it is further suppressed by squaring it, this term can be neglected.
The integration region $\mathcal{R}(t)$ is defined to be a low-frequency range that allows us to neglect the UV divergence issue, which can be chosen the same as our early time approximation condition $\omega_{\mathbf{k}}t < 1$.
We first define $k_{yz} = \sqrt{k_y^2+k_z^2}$, and set $a(t_0)=1$. The domain of integration $\mathcal{R}(t)$ is an ellipse in $k$ space:
\begin{align}
    (k_{yz}t)^2+(k_xt/a)^2 \leq 1\\
    \Rightarrow k_{yz}^2+\qty(\frac{k_x}{a})^2 \leq \frac{1}{t^2}.
\end{align}
Recognizing the axial symmetry in the problem, we employ cylindrical coordinates and find
\begin{align}
\rho&=(576 \pi ^2 a^{10/3} t^4)^{-1}\Biggl[ \Biggr.9 a^4-36 a^{10/3}+18 a^{8/3} \mathcal{P} +9 a^2 \mathcal{P}-4 a'^2 \mathcal{P} t^3\Biggl.\Biggr] + \rho_{\textsc{CA}},\\
&= \rho_{d}+\rho_{\textsc{CA}}
\end{align}
where 
\[\mathcal{P}=
\begin{cases}
 \frac{\sin ^{-1}\left(\sqrt{a^{-2}-1} a\right)}{\sqrt{a^{-2}-1}} & a<1,\\
 \frac{\log \left(\left(\sqrt{1-a^{-2}}+1\right) a\right)}{\sqrt{1-a^{-2}}} & a>1, \\
 1 & a=1.
\end{cases}\]

\begin{figure}[htbp]
    \centering
    \includegraphics[width=0.45\columnwidth]{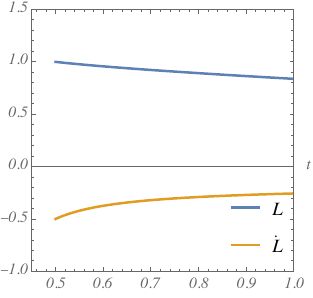}
    \includegraphics[width=0.45\columnwidth]{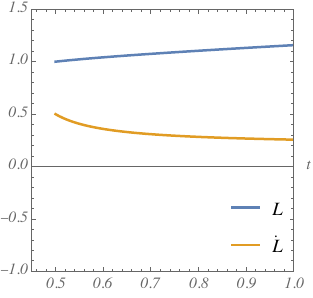}
    \caption{Plot of mirror's position and velocity vs. time. Left figure for $\dot{L}(t_0)=-0.5$, right figure for $\dot{L}(t_0)=0.5$.}
    \label{fig:4d dy}
\end{figure}

We can construct an action for the dynamics of this moving mirror system by placing the energy without Casimir energy as the kinetic terms, and $\rho_\text{CA}$ as a potential term:
\begin{align}
    S&= \int \dd t \qty{\frac{1}{2}m\dot{L}^2 + \int\displaylimits_{\text{box}} d^3x \qty[\rho_{d}-\rho_{\text{CA}}]}\\
    &= \int \dd t \qty{\frac{1}{2}m\dot{L}^2 + Ll^2 \qty[\rho_{d}-\rho_{\text{CA}}]}\\
    &= \int \dd t \qty{\frac{1}{2}m\dot{L}^2 + E_{d}-E_{\text{CA}}}
\end{align}
where $L \equiv al$ is the time-dependent longitudinal length of the box.
The Casimir energy Eq. \eqref{4dcas} is quite complicated and does not allow a quick numerical evaluation. However, its behavior is quite simple to understand, just contributes an attractive force to the moving mirror. We do not consider the effect of the Casimir energy for now. 
\begin{equation}
    S= \int \dd t \qty{\frac{1}{2}m\dot{L}^2 + E_\text{d}}
    \label{box action}
\end{equation}
Assuming that the moving mirror is a classical object, taking the functional variation of the action in Eq. \eqref{box action} gives the Euler-Lagrange equation for the dynamics of the mirror. We do not provide the explicit expression for this equation, since it is particularly complicated. It is numerically solved for cases when the mirror initially is left moving (expanding box) and ii) when the mirror is initially right moving (shrinking box). We chose parameters $l=50$, $m=10$, and $t_0 = 0.5$\footnote{Note the choice of $t_0=0.5$ is simply avoid the $t=0$ singularity, caused by our choice of integration range, this does not have any physical meaning, and the choice of integration range is quite flexible. }.  The results are shown in Fig. \ref{fig:4d dy} 

It is shown that the mirror slows down in the same way no matter the direction it initially moving. 

\newpage
\bibliography{refs}
\end{document}